\documentclass{elsart}

\usepackage{amssymb}

\begin{document}

\begin{frontmatter}



\title{A Fortran Code for Null Geodesic Solutions in the
       Lema\^{\i}tre-Tolman-Bondi Spacetime}


\author{Marcelo B.\ Ribeiro\thanksref{if}\thanksref{capes}}

\address{Vatican Observatory Group, Steward Observatory, University of
         Arizona, Tucson, AZ 85721, USA; e-mail: mribeiro@as.arizona.edu} 

\thanks[if]{On leave from Physics Institute, University of Brazil -- UFRJ,
            Rio de Janeiro.}
\thanks[capes]{Supported by Brazil's CAPES Foundation.}

\begin{abstract}
This paper describes the Fortran 77 code {\tt SIMU}, version 1.1, designed
for numerical simulations of observational relations along the past null
geodesic in the Lema\^{\i}tre-Tolman-Bondi (LTB) spacetime. {\tt SIMU}
aims at finding scale invariant solutions of the average density, but due
to its full modularity it can be easily adapted to any application which
requires LTB's null geodesic solutions. In version 1.1 the numerical
output can be read by the {\tt GNUPLOT} plotting package to produce a
fully graphical output, although other plotting routines can be easily
adapted. Details of the code's subroutines are discussed, and an example
of its output is shown.
\end{abstract}

\end{frontmatter}

% main text
\section{Introduction}\label{int}

It is a basic result of relativity theory that light rays follow null
geodesics in curved 4-dimensional spacetimes, and this means that if one
is pursuing cosmological applications of solutions of Einstein's field
equations with the aim of comparing the model's theoretical predictions
with the actual data produced by astronomical observations, one requires
null geodesic solutions for the chosen metric. In fact, for cosmological
applications one only needs the {\it past} null geodesics, as what we
observe today are events which occurred in the past. Nevertheless, a
very basic problem arising in cosmology, and which hinders comparison
between theory and observations in other models than the standard
Friedmann-Lema\^{\i}tre-Robertson-Walker (FLRW), stems from the fact
that as soon as we depart from the simple FLRW models largely used by
observational cosmologists, the task of finding null geodesic solutions
quickly becomes an intractable analytical problem. In fact, this
situation is not fully appreciated by many of those dealing with
cosmological modelling of astronomical data, inasmuch as the simple
FLRW models largely applied in observational cosmology is the exception,
since it has a very simple analytical solution for the null geodesic
equation.

The Lema\^{\i}tre-Tolman-Bondi (LTB) spacetime is the most general
spherically symmetric dust solution of Einstein's field equations,
having the FLRW models as special sub-cases \cite{rib92b,rib94,kr}. By being
a spatially inhomogeneous model, that is, where $\rho=\rho(r,t)$, it allows
us to study scenarios which do not have FLRW's spatial homogeneity assumed
{\it a priori}. Due to this, it is the most widely used cosmological model
after the standard FLRW, having been applied in a wide range of cosmological
problems, from microwave background radiation studies to hierarchical
(fractal) modelling, just to quote a few (see \cite{kr}, and references
therein, for a large number of applications of LTB models in cosmology).
However, as is the case with almost all non-standard cosmologies, while
it is possible to solve analytically its Einstein's field equations,
LTB's null geodesic equation remains an intractable analytical problem,
unless we rewrite its geometry in terms of the so-called ``observational
coordinates''. That, however, has the handicap of adding a great deal of
mathematical complexity to the problem, since such an approach requires
complex mathematical calculations in curved spacetimes \cite{mhms},
something which may not be required in all LTB applications. Therefore,
due to such a wide range of application, a code which produces numerical
solutions of LTB null geodesics is desirable.

In this paper I will describe such a code. It deals with LTB geometry in
its full generality, being applicable to each of its special sub-cases,
parabolic, hyperbolic and elliptic \cite{bonnor74}, either separately or
together in a single problem encompassing all sub-cases, if desired. It
was originally designed to find scale invariant solutions of the average
density by means of numerical simulations \cite{rib92,rib93}, but as we
shall see below, without changing its core null geodesic calculations the
code can be easily adapted to do much more than this, due to its
modularity. The original results obtained by this code were recently
analytically confirmed, and extended, by other authors \cite{matra} by
means of the observational coordinates approach mentioned above. The
confirmation of the results of \cite{rib93} by \cite{matra} adds then
further reliability to the numerical results obtained with the code that
will be described below. Besides, those two approaches, numerical and
analytical, when applied to a difficult problem such as null geodesic
solutions in LTB spacetime, will in fact complement each other, rather
than exclude one another. In \S \ref{des} I will describe version 1.1 of
the code, and in \S \ref{io} the input and output will be discussed. The
results are summarized in \S \ref{conc}, where I also indicate where the
code can be obtained.

\section{Description of {\tt SIMU} 1.1}\label{des}

{\tt SIMU} solves simultaneously two ordinary differential equations
(ODE's) by means of the fourth order Runge-Kutta method with adaptive
stepsize control. One differential equation is required for the null
geodesic, while the other is needed for obtaining the redshift in this
cosmology \cite{rib92}. The subroutines that carry out the numerical
integration are from \cite{nr}, and some minor changes necessary for
taking the results back to the main program were made, but without
changing the actual numerical implementation of the Runge-Kutta method.
A full description of the LTB model and notation as used in {\tt SIMU}
is given in \cite{rib92}. The numerical simulations take advantage of
the fact that LTB spacetime geometry has three unknown
functions,\footnote{ \ Actually two, as a third  can be eliminated by
a coordinate transformation.}$F(r)$, $f(r)$, $\beta(r)$, respectively
representing the amount of gravitational mass within the radius coordinate
$r$, the overall curvature and dynamics of the model, and the time elapsed
since the big bang for each observer located at particular values of $r$
\cite{rib94,kr,bonnor74}. The simulations then take advantage of this
freedom, as one starts by choosing these functions, run the program
and analyse the results, concluding then whether or not the simulation
was successful, and if not choose another set of three functions (see
\cite{rib93} for a very detailed explanation of this procedure).

The code is implemented in double precision, and it has a built-in
methodology for checking the possible catastrophic accumulation of
round-off errors. That is done by monitoring the energy equation, as
derived from Einstein's field equations, and its derivative, since, by
theory, both must remain unchanged throughout the integration
\cite{rib93}. In addition, to check for accuracy and stability, after
the integration the program runs in reverse, that is, it takes the
final result and uses it as initial conditions to get another result
that can be compared with the original initial condition. The results
described in \cite{rib93} showed that double precision is enough for
maintaining the desired accuracy required by the problem under study.
{\tt SIMU} also calculates the errors associated with each observational
quantity evaluated, as well as the propagated errors. This is done by
taking the accuracy given by the adaptive method and using this value
as the input in the standard propagation error equations. All those
details are fully explained in the initial documentation of the code.

As mentioned above, the code uses some subroutines from \cite{nr},
namely the fourth order Runge-Kutta integration with adaptive stepsize
control, the root finding algorithm for the transcendental equations
appearing in LTB's elliptical and hyperbolic sub-cases \cite{rib92},
the extended trapezoidal rule quadrature algorithm, as well as Simpson's
rule to the desired accuracy. However, apart from those, the remaining
subroutines are new, as well as the actual way in which the whole set of
subroutines were linked to each other and organized. In addition, the
code is extensively commented and widely documented, also having a
structure chart showing the dependence among the subroutines and
functions. This detailed documentation is added in order to help
readers who might be interested in implementing {\tt SIMU} in their
own computer environments, or changing it to suit their specific
applications.

It must be mentioned that there are more ``state of the art'' ways
for finding numerical solutions of ODE's than using the adaptive
4th order Runge-Kutta method. The ODEPACK subroutine package is an
example \cite{r2,r3,r4}. Nevertheless, although the chosen ODE integrator
for {\tt SIMU} may have already been superseded by better methods, it
{\it does} solve the proposed problem, and at the desired accuracy.
Inasmuch as {\tt SIMU} is a tested code, with its results having already
been analytically confirmed by other authors \cite{matra}, there is no
need to change to another ODE integrator at {\tt SIMU}'s current version.
The reader must, however, be aware that such a change might be required
for other applications of {\tt SIMU} to LTB spacetime.

The code uses only standard Fortran 77 commands to avoid possible
compilation problems with machine dependent commands that are not
available universally. If a potential user is only interested in the
core of the program, that is, the part which integrates the null
geodesic and calculates the observational relations, he or she can
simply remove its ``tail'', where the observational relations are
manipulated according to the aims of the theory studied in
\cite{rib92,rib93}, and replace it with something else. This ``tail''
consists of the two last branches of the structure chart headed by
the subroutines {\tt FIT} and {\tt OUTPUT}, and are easily spotted
in the {\tt MAIN}. Finally, the functions $f(r)$, $F(r)$, $\beta(r)$
and their derivatives appear in six subroutines at the end of the
list so that they can be compiled separately.

The previous version 1.0 (formerly {\tt SIMU 5d}; see \cite{thesis}) had
the {\tt PGPLOT} plotting routines merged into the second half of the
{\tt OUTPUT} subroutine, while version 1.1 removed them and piped the
numerical results into eleven independent files, {\tt v1} to {\tt v11},
that are then {\it externally} read by the {\tt GNUPLOT} plotting
package in order to produce a graphical output of the results. This is
the only difference between versions 1.0 and 1.1, meaning that in
version 1.1 the graphical output is completely independent from the
code itself. Making this change was justified on two grounds: {\it (i)}
the {\tt GNUPLOT} package has greater plotting capabilities, even
allowing \LaTeX \ output formats that can be included directly in the
{\tt figure} environment of any \LaTeX \ macro. So, producing plots
with the results is done by feeding the numerical tables into {\tt
GNUPLOT} itself by means of a script file written for, say, a linux
environment; {\it (ii)} this feature of having the results outputted
into numerical files brings flexibility to the code, as its results
can be independently used by another program, if so desired or demanded
by another application of the LTB spacetime.

\section{Input and Output}\label{io}

The actual implementation for the original problem advanced in
\cite{rib92} was described at length in \cite{rib93}, where one can find
detailed plots with the results and analysis of the various
simulations,\footnote{ \ These results were summarized in \cite{rib94}.}
as well as detailed explanations of initial conditions and functions used
in each simulation. So, in here I shall limit the discussion to the code
itself, without dealing with any specific application. 

As mentioned above, we require a script file for running a
simulation and immediately producing a graphical output. An example of
such a script in linux is given as follows:

{\footnotesize \tt g77 simu1.1.for\\a.out < in.simu\\gnuplot < in.gnu\\latex
plots.tex\\xdvi plots.dvi}

Here the file {\tt in.simu} contains the maximum value for the radius
coordinate $r$ for the integration, while {\tt in.gnu} is a short
{\tt GNUPLOT} script file for inputting the {\tt v1} to {\tt v11} files
and producing \LaTeX \ files (see appendix \ref{app1}). Finally
{\tt plots.tex} is a file for running the {\tt .tex} \LaTeX \ files produced
by {\tt GNUPLOT} in order for the results be available for analysis
directly on the screen.

The details of each simulations, as well as accuracy monitoring are given
in an outputted file called {\tt s1}. Appendix \ref{app2} shows the
results of a simulation using the value 1.5 for {\tt in.simu}, and the
three functions given as follows: $f(r)=\cosh(r)$, $F(r)=\sinh^3(r)$,
$\beta(r)=0.7$. 

\section{Conclusion}\label{conc}

In this paper I have described the Fortran 77 code {\tt SIMU} version
1.1 for calculating solutions of the null geodesics equations in the
Lema\^{\i}tre-Tolman-Bondi spacetime geometry. I have discussed the
details of the code, input and output, as well as the ways in which it
can be possibly modified for other cosmological applications of this
geometry. The code is available for download at \cite{distro}.

\ack

I am grateful to William R.\ Stoeger for reading the original manuscript,
Newton Jos\'e de Moura Jr.\ for his assistance with the {\tt GNUPLOT}
package, and the referee for helpful comments. I also thank the Vatican
Observatory Research Group for their kind hospitality while writing this
paper.

\appendix
\section{{\tt in.gnu} Script}\label{app1}

{\footnotesize \tt
set terminal latex\\
set output 'v1.tex'\\
set size 1,1\\  
set xlabel "\$ $\backslash$log d\_l  (Gpc)\$"\\
set ylabel "\$ $\backslash$log  $\backslash$rho\$"\\
set nokey\\
plot 'v1' with points 1 8, 'v11' with lines 1\\ 
set output 'v2.tex'\\
set xlabel "\$d\_l  (Gpc)\$"\\
set ylabel "\$ $\backslash$rho\$"\\
plot 'v2' with points 1 8\\
set output 'v3.tex'\\
set xlabel "\$z\$"\\
set ylabel "\$d\_l\$ $\backslash$$\backslash$ \$(Gpc)\$"\\
plot 'v3' with points 1 8\\
set output 'v4.tex'\\
set xlabel "\$z\$"\\
set ylabel "\$N\_c\$"\\
plot 'v4' with points 1 8\\
set output 'v5.tex'\\
set xlabel "\$r\$"\\
set ylabel "\$Field\$ $\backslash$$\backslash$ \$EE\$"\\
set yrange [-10**-5:10**-5]\\
plot 'v5' with points 1 8\\
set output 'v6.tex'\\
set xlabel "\$r\$"\\
set ylabel "\$Deriv.\$ $\backslash$$\backslash$ \$Field\$
$\backslash$$\backslash$ \$EEDSH\$"\\
set yrange [-10**-5:10**-5]\\
plot 'v6' with points 1 8\\
set output 'v7.tex'\\
set title "\$Center  $\backslash$rightarrow Border\$"\\
set xlabel "\$r\$"\\
set ylabel "\$t\$"\\
set autoscale y\\
plot 'v7' with points 1 8\\
set output 'v8.tex'\\
set title "\$Border  $\backslash$rightarrow Center\$"\\
set xlabel "\$r\$"\\
set ylabel "\$t\$"\\
plot 'v8' with points 1 8\\
set output 'v9.tex'\\
set title "\$Center  $\backslash$rightarrow Border\$"\\
set xlabel "\$r\$"\\
set ylabel "\$I\$"\\
plot 'v9' with points 1 8\\
set output 'v10.tex'\\
set title "\$Border  $\backslash$rightarrow Center\$"\\
set xlabel "\$r\$"\\
set ylabel "\$I\$"\\
plot 'v10' with points 1 8\\
quit
}

\newpage
\section{Plots of a Single Simulation}\label{app2}

\begin{figure}[h]
    \input{v1}
\end{figure}
\begin{figure}[b]
    % GNUPLOT: LaTeX picture
\setlength{\unitlength}{0.240900pt}
\ifx\plotpoint\undefined\newsavebox{\plotpoint}\fi
\begin{picture}(1500,900)(0,0)
\sbox{\plotpoint}{\rule[-0.175pt]{0.350pt}{0.350pt}}%
\put(264,158){\rule[-0.175pt]{282.335pt}{0.350pt}}
\put(264,158){\rule[-0.175pt]{0.350pt}{151.526pt}}
\put(264,158){\rule[-0.175pt]{4.818pt}{0.350pt}}
\put(242,158){\makebox(0,0)[r]{0}}
\put(1416,158){\rule[-0.175pt]{4.818pt}{0.350pt}}
\put(264,237){\rule[-0.175pt]{4.818pt}{0.350pt}}
\put(242,237){\makebox(0,0)[r]{0.0001}}
\put(1416,237){\rule[-0.175pt]{4.818pt}{0.350pt}}
\put(264,315){\rule[-0.175pt]{4.818pt}{0.350pt}}
\put(242,315){\makebox(0,0)[r]{0.0002}}
\put(1416,315){\rule[-0.175pt]{4.818pt}{0.350pt}}
\put(264,394){\rule[-0.175pt]{4.818pt}{0.350pt}}
\put(242,394){\makebox(0,0)[r]{0.0003}}
\put(1416,394){\rule[-0.175pt]{4.818pt}{0.350pt}}
\put(264,473){\rule[-0.175pt]{4.818pt}{0.350pt}}
\put(242,473){\makebox(0,0)[r]{0.0004}}
\put(1416,473){\rule[-0.175pt]{4.818pt}{0.350pt}}
\put(264,551){\rule[-0.175pt]{4.818pt}{0.350pt}}
\put(242,551){\makebox(0,0)[r]{0.0005}}
\put(1416,551){\rule[-0.175pt]{4.818pt}{0.350pt}}
\put(264,630){\rule[-0.175pt]{4.818pt}{0.350pt}}
\put(242,630){\makebox(0,0)[r]{0.0006}}
\put(1416,630){\rule[-0.175pt]{4.818pt}{0.350pt}}
\put(264,708){\rule[-0.175pt]{4.818pt}{0.350pt}}
\put(242,708){\makebox(0,0)[r]{0.0007}}
\put(1416,708){\rule[-0.175pt]{4.818pt}{0.350pt}}
\put(264,787){\rule[-0.175pt]{4.818pt}{0.350pt}}
\put(242,787){\makebox(0,0)[r]{0.0008}}
\put(1416,787){\rule[-0.175pt]{4.818pt}{0.350pt}}
\put(264,158){\rule[-0.175pt]{0.350pt}{4.818pt}}
\put(264,113){\makebox(0,0){0}}
\put(264,767){\rule[-0.175pt]{0.350pt}{4.818pt}}
\put(459,158){\rule[-0.175pt]{0.350pt}{4.818pt}}
\put(459,113){\makebox(0,0){10}}
\put(459,767){\rule[-0.175pt]{0.350pt}{4.818pt}}
\put(655,158){\rule[-0.175pt]{0.350pt}{4.818pt}}
\put(655,113){\makebox(0,0){20}}
\put(655,767){\rule[-0.175pt]{0.350pt}{4.818pt}}
\put(850,158){\rule[-0.175pt]{0.350pt}{4.818pt}}
\put(850,113){\makebox(0,0){30}}
\put(850,767){\rule[-0.175pt]{0.350pt}{4.818pt}}
\put(1045,158){\rule[-0.175pt]{0.350pt}{4.818pt}}
\put(1045,113){\makebox(0,0){40}}
\put(1045,767){\rule[-0.175pt]{0.350pt}{4.818pt}}
\put(1241,158){\rule[-0.175pt]{0.350pt}{4.818pt}}
\put(1241,113){\makebox(0,0){50}}
\put(1241,767){\rule[-0.175pt]{0.350pt}{4.818pt}}
\put(1436,158){\rule[-0.175pt]{0.350pt}{4.818pt}}
\put(1436,113){\makebox(0,0){60}}
\put(1436,767){\rule[-0.175pt]{0.350pt}{4.818pt}}
\put(264,158){\rule[-0.175pt]{282.335pt}{0.350pt}}
\put(1436,158){\rule[-0.175pt]{0.350pt}{151.526pt}}
\put(264,787){\rule[-0.175pt]{282.335pt}{0.350pt}}
\put(45,472){\makebox(0,0)[l]{\shortstack{$\rho$}}}
\put(850,68){\makebox(0,0){$d_l  (Gpc)$}}
\put(264,158){\rule[-0.175pt]{0.350pt}{151.526pt}}
\put(264,158){\circle{18}}
\put(264,756){\circle{18}}
\put(266,724){\circle{18}}
\put(272,622){\circle{18}}
\put(287,457){\circle{18}}
\put(315,323){\circle{18}}
\put(359,244){\circle{18}}
\put(434,197){\circle{18}}
\put(557,174){\circle{18}}
\put(765,166){\circle{18}}
\put(1133,158){\circle{18}}
\put(1242,158){\circle{18}}
\end{picture}
\end{figure}
\begin{figure}[htb]
    % GNUPLOT: LaTeX picture
\setlength{\unitlength}{0.240900pt}
\ifx\plotpoint\undefined\newsavebox{\plotpoint}\fi
\begin{picture}(1500,900)(0,0)
\sbox{\plotpoint}{\rule[-0.175pt]{0.350pt}{0.350pt}}%
\put(264,158){\rule[-0.175pt]{282.335pt}{0.350pt}}
\put(264,158){\rule[-0.175pt]{0.350pt}{151.526pt}}
\put(264,158){\rule[-0.175pt]{4.818pt}{0.350pt}}
\put(242,158){\makebox(0,0)[r]{0}}
\put(1416,158){\rule[-0.175pt]{4.818pt}{0.350pt}}
\put(264,263){\rule[-0.175pt]{4.818pt}{0.350pt}}
\put(242,263){\makebox(0,0)[r]{10}}
\put(1416,263){\rule[-0.175pt]{4.818pt}{0.350pt}}
\put(264,368){\rule[-0.175pt]{4.818pt}{0.350pt}}
\put(242,368){\makebox(0,0)[r]{20}}
\put(1416,368){\rule[-0.175pt]{4.818pt}{0.350pt}}
\put(264,473){\rule[-0.175pt]{4.818pt}{0.350pt}}
\put(242,473){\makebox(0,0)[r]{30}}
\put(1416,473){\rule[-0.175pt]{4.818pt}{0.350pt}}
\put(264,577){\rule[-0.175pt]{4.818pt}{0.350pt}}
\put(242,577){\makebox(0,0)[r]{40}}
\put(1416,577){\rule[-0.175pt]{4.818pt}{0.350pt}}
\put(264,682){\rule[-0.175pt]{4.818pt}{0.350pt}}
\put(242,682){\makebox(0,0)[r]{50}}
\put(1416,682){\rule[-0.175pt]{4.818pt}{0.350pt}}
\put(264,787){\rule[-0.175pt]{4.818pt}{0.350pt}}
\put(242,787){\makebox(0,0)[r]{60}}
\put(1416,787){\rule[-0.175pt]{4.818pt}{0.350pt}}
\put(264,158){\rule[-0.175pt]{0.350pt}{4.818pt}}
\put(264,113){\makebox(0,0){0}}
\put(264,767){\rule[-0.175pt]{0.350pt}{4.818pt}}
\put(381,158){\rule[-0.175pt]{0.350pt}{4.818pt}}
\put(381,113){\makebox(0,0){0.5}}
\put(381,767){\rule[-0.175pt]{0.350pt}{4.818pt}}
\put(498,158){\rule[-0.175pt]{0.350pt}{4.818pt}}
\put(498,113){\makebox(0,0){1}}
\put(498,767){\rule[-0.175pt]{0.350pt}{4.818pt}}
\put(616,158){\rule[-0.175pt]{0.350pt}{4.818pt}}
\put(616,113){\makebox(0,0){1.5}}
\put(616,767){\rule[-0.175pt]{0.350pt}{4.818pt}}
\put(733,158){\rule[-0.175pt]{0.350pt}{4.818pt}}
\put(733,113){\makebox(0,0){2}}
\put(733,767){\rule[-0.175pt]{0.350pt}{4.818pt}}
\put(850,158){\rule[-0.175pt]{0.350pt}{4.818pt}}
\put(850,113){\makebox(0,0){2.5}}
\put(850,767){\rule[-0.175pt]{0.350pt}{4.818pt}}
\put(967,158){\rule[-0.175pt]{0.350pt}{4.818pt}}
\put(967,113){\makebox(0,0){3}}
\put(967,767){\rule[-0.175pt]{0.350pt}{4.818pt}}
\put(1084,158){\rule[-0.175pt]{0.350pt}{4.818pt}}
\put(1084,113){\makebox(0,0){3.5}}
\put(1084,767){\rule[-0.175pt]{0.350pt}{4.818pt}}
\put(1202,158){\rule[-0.175pt]{0.350pt}{4.818pt}}
\put(1202,113){\makebox(0,0){4}}
\put(1202,767){\rule[-0.175pt]{0.350pt}{4.818pt}}
\put(1319,158){\rule[-0.175pt]{0.350pt}{4.818pt}}
\put(1319,113){\makebox(0,0){4.5}}
\put(1319,767){\rule[-0.175pt]{0.350pt}{4.818pt}}
\put(1436,158){\rule[-0.175pt]{0.350pt}{4.818pt}}
\put(1436,113){\makebox(0,0){5}}
\put(1436,767){\rule[-0.175pt]{0.350pt}{4.818pt}}
\put(264,158){\rule[-0.175pt]{282.335pt}{0.350pt}}
\put(1436,158){\rule[-0.175pt]{0.350pt}{151.526pt}}
\put(264,787){\rule[-0.175pt]{282.335pt}{0.350pt}}
\put(45,472){\makebox(0,0)[l]{\shortstack{$d_l$\\$(Gpc)$}}}
\put(850,68){\makebox(0,0){$z$}}
\put(264,158){\rule[-0.175pt]{0.350pt}{151.526pt}}
\put(264,158){\circle{18}}
\put(265,158){\circle{18}}
\put(270,159){\circle{18}}
\put(286,162){\circle{18}}
\put(326,170){\circle{18}}
\put(388,186){\circle{18}}
\put(468,209){\circle{18}}
\put(581,249){\circle{18}}
\put(733,315){\circle{18}}
\put(944,427){\circle{18}}
\put(1246,624){\circle{18}}
\put(1325,683){\circle{18}}
\end{picture}
\end{figure}
\begin{figure}[htb]
    % GNUPLOT: LaTeX picture
\setlength{\unitlength}{0.240900pt}
\ifx\plotpoint\undefined\newsavebox{\plotpoint}\fi
\begin{picture}(1500,900)(0,0)
\sbox{\plotpoint}{\rule[-0.175pt]{0.350pt}{0.350pt}}%
\put(264,158){\rule[-0.175pt]{282.335pt}{0.350pt}}
\put(264,158){\rule[-0.175pt]{0.350pt}{151.526pt}}
\put(264,158){\rule[-0.175pt]{4.818pt}{0.350pt}}
\put(242,158){\makebox(0,0)[r]{0}}
\put(1416,158){\rule[-0.175pt]{4.818pt}{0.350pt}}
\put(264,263){\rule[-0.175pt]{4.818pt}{0.350pt}}
\put(242,263){\makebox(0,0)[r]{5e+10}}
\put(1416,263){\rule[-0.175pt]{4.818pt}{0.350pt}}
\put(264,368){\rule[-0.175pt]{4.818pt}{0.350pt}}
\put(242,368){\makebox(0,0)[r]{1e+11}}
\put(1416,368){\rule[-0.175pt]{4.818pt}{0.350pt}}
\put(264,473){\rule[-0.175pt]{4.818pt}{0.350pt}}
\put(242,473){\makebox(0,0)[r]{1.5e+11}}
\put(1416,473){\rule[-0.175pt]{4.818pt}{0.350pt}}
\put(264,577){\rule[-0.175pt]{4.818pt}{0.350pt}}
\put(242,577){\makebox(0,0)[r]{2e+11}}
\put(1416,577){\rule[-0.175pt]{4.818pt}{0.350pt}}
\put(264,682){\rule[-0.175pt]{4.818pt}{0.350pt}}
\put(242,682){\makebox(0,0)[r]{2.5e+11}}
\put(1416,682){\rule[-0.175pt]{4.818pt}{0.350pt}}
\put(264,787){\rule[-0.175pt]{4.818pt}{0.350pt}}
\put(242,787){\makebox(0,0)[r]{3e+11}}
\put(1416,787){\rule[-0.175pt]{4.818pt}{0.350pt}}
\put(264,158){\rule[-0.175pt]{0.350pt}{4.818pt}}
\put(264,113){\makebox(0,0){0}}
\put(264,767){\rule[-0.175pt]{0.350pt}{4.818pt}}
\put(381,158){\rule[-0.175pt]{0.350pt}{4.818pt}}
\put(381,113){\makebox(0,0){0.5}}
\put(381,767){\rule[-0.175pt]{0.350pt}{4.818pt}}
\put(498,158){\rule[-0.175pt]{0.350pt}{4.818pt}}
\put(498,113){\makebox(0,0){1}}
\put(498,767){\rule[-0.175pt]{0.350pt}{4.818pt}}
\put(616,158){\rule[-0.175pt]{0.350pt}{4.818pt}}
\put(616,113){\makebox(0,0){1.5}}
\put(616,767){\rule[-0.175pt]{0.350pt}{4.818pt}}
\put(733,158){\rule[-0.175pt]{0.350pt}{4.818pt}}
\put(733,113){\makebox(0,0){2}}
\put(733,767){\rule[-0.175pt]{0.350pt}{4.818pt}}
\put(850,158){\rule[-0.175pt]{0.350pt}{4.818pt}}
\put(850,113){\makebox(0,0){2.5}}
\put(850,767){\rule[-0.175pt]{0.350pt}{4.818pt}}
\put(967,158){\rule[-0.175pt]{0.350pt}{4.818pt}}
\put(967,113){\makebox(0,0){3}}
\put(967,767){\rule[-0.175pt]{0.350pt}{4.818pt}}
\put(1084,158){\rule[-0.175pt]{0.350pt}{4.818pt}}
\put(1084,113){\makebox(0,0){3.5}}
\put(1084,767){\rule[-0.175pt]{0.350pt}{4.818pt}}
\put(1202,158){\rule[-0.175pt]{0.350pt}{4.818pt}}
\put(1202,113){\makebox(0,0){4}}
\put(1202,767){\rule[-0.175pt]{0.350pt}{4.818pt}}
\put(1319,158){\rule[-0.175pt]{0.350pt}{4.818pt}}
\put(1319,113){\makebox(0,0){4.5}}
\put(1319,767){\rule[-0.175pt]{0.350pt}{4.818pt}}
\put(1436,158){\rule[-0.175pt]{0.350pt}{4.818pt}}
\put(1436,113){\makebox(0,0){5}}
\put(1436,767){\rule[-0.175pt]{0.350pt}{4.818pt}}
\put(264,158){\rule[-0.175pt]{282.335pt}{0.350pt}}
\put(1436,158){\rule[-0.175pt]{0.350pt}{151.526pt}}
\put(264,787){\rule[-0.175pt]{282.335pt}{0.350pt}}
\put(45,472){\makebox(0,0)[l]{\shortstack{$N_c$}}}
\put(850,68){\makebox(0,0){$z$}}
\put(264,158){\rule[-0.175pt]{0.350pt}{151.526pt}}
\put(264,158){\circle{18}}
\put(265,158){\circle{18}}
\put(270,158){\circle{18}}
\put(286,158){\circle{18}}
\put(326,159){\circle{18}}
\put(388,165){\circle{18}}
\put(468,181){\circle{18}}
\put(581,218){\circle{18}}
\put(733,291){\circle{18}}
\put(944,425){\circle{18}}
\put(1246,664){\circle{18}}
\put(1325,734){\circle{18}}
\end{picture}
\end{figure}
\begin{figure}[htb]
    % GNUPLOT: LaTeX picture
\setlength{\unitlength}{0.240900pt}
\ifx\plotpoint\undefined\newsavebox{\plotpoint}\fi
\begin{picture}(1500,900)(0,0)
\sbox{\plotpoint}{\rule[-0.175pt]{0.350pt}{0.350pt}}%
\put(264,473){\rule[-0.175pt]{282.335pt}{0.350pt}}
\put(264,158){\rule[-0.175pt]{0.350pt}{151.526pt}}
\put(264,158){\rule[-0.175pt]{4.818pt}{0.350pt}}
\put(242,158){\makebox(0,0)[r]{-1e-05}}
\put(1416,158){\rule[-0.175pt]{4.818pt}{0.350pt}}
\put(264,315){\rule[-0.175pt]{4.818pt}{0.350pt}}
\put(242,315){\makebox(0,0)[r]{-5e-06}}
\put(1416,315){\rule[-0.175pt]{4.818pt}{0.350pt}}
\put(264,473){\rule[-0.175pt]{4.818pt}{0.350pt}}
\put(242,473){\makebox(0,0)[r]{0}}
\put(1416,473){\rule[-0.175pt]{4.818pt}{0.350pt}}
\put(264,630){\rule[-0.175pt]{4.818pt}{0.350pt}}
\put(242,630){\makebox(0,0)[r]{5e-06}}
\put(1416,630){\rule[-0.175pt]{4.818pt}{0.350pt}}
\put(264,787){\rule[-0.175pt]{4.818pt}{0.350pt}}
\put(242,787){\makebox(0,0)[r]{1e-05}}
\put(1416,787){\rule[-0.175pt]{4.818pt}{0.350pt}}
\put(264,158){\rule[-0.175pt]{0.350pt}{4.818pt}}
\put(264,113){\makebox(0,0){0}}
\put(264,767){\rule[-0.175pt]{0.350pt}{4.818pt}}
\put(411,158){\rule[-0.175pt]{0.350pt}{4.818pt}}
\put(411,113){\makebox(0,0){0.2}}
\put(411,767){\rule[-0.175pt]{0.350pt}{4.818pt}}
\put(557,158){\rule[-0.175pt]{0.350pt}{4.818pt}}
\put(557,113){\makebox(0,0){0.4}}
\put(557,767){\rule[-0.175pt]{0.350pt}{4.818pt}}
\put(704,158){\rule[-0.175pt]{0.350pt}{4.818pt}}
\put(704,113){\makebox(0,0){0.6}}
\put(704,767){\rule[-0.175pt]{0.350pt}{4.818pt}}
\put(850,158){\rule[-0.175pt]{0.350pt}{4.818pt}}
\put(850,113){\makebox(0,0){0.8}}
\put(850,767){\rule[-0.175pt]{0.350pt}{4.818pt}}
\put(997,158){\rule[-0.175pt]{0.350pt}{4.818pt}}
\put(997,113){\makebox(0,0){1}}
\put(997,767){\rule[-0.175pt]{0.350pt}{4.818pt}}
\put(1143,158){\rule[-0.175pt]{0.350pt}{4.818pt}}
\put(1143,113){\makebox(0,0){1.2}}
\put(1143,767){\rule[-0.175pt]{0.350pt}{4.818pt}}
\put(1290,158){\rule[-0.175pt]{0.350pt}{4.818pt}}
\put(1290,113){\makebox(0,0){1.4}}
\put(1290,767){\rule[-0.175pt]{0.350pt}{4.818pt}}
\put(1436,158){\rule[-0.175pt]{0.350pt}{4.818pt}}
\put(1436,113){\makebox(0,0){1.6}}
\put(1436,767){\rule[-0.175pt]{0.350pt}{4.818pt}}
\put(264,158){\rule[-0.175pt]{282.335pt}{0.350pt}}
\put(1436,158){\rule[-0.175pt]{0.350pt}{151.526pt}}
\put(264,787){\rule[-0.175pt]{282.335pt}{0.350pt}}
\put(45,472){\makebox(0,0)[l]{\shortstack{$Field$\\$EE$}}}
\put(850,68){\makebox(0,0){$r$}}
\put(264,158){\rule[-0.175pt]{0.350pt}{151.526pt}}
\put(264,473){\circle{18}}
\put(268,473){\circle{18}}
\put(281,473){\circle{18}}
\put(325,472){\circle{18}}
\put(425,473){\circle{18}}
\put(554,472){\circle{18}}
\put(689,472){\circle{18}}
\put(840,473){\circle{18}}
\put(996,473){\circle{18}}
\put(1158,472){\circle{18}}
\put(1327,473){\circle{18}}
\put(1363,473){\circle{18}}
\end{picture}
\end{figure}
\begin{figure}[htb]
    % GNUPLOT: LaTeX picture
\setlength{\unitlength}{0.240900pt}
\ifx\plotpoint\undefined\newsavebox{\plotpoint}\fi
\begin{picture}(1500,900)(0,0)
\sbox{\plotpoint}{\rule[-0.175pt]{0.350pt}{0.350pt}}%
\put(264,473){\rule[-0.175pt]{282.335pt}{0.350pt}}
\put(264,158){\rule[-0.175pt]{0.350pt}{151.526pt}}
\put(264,158){\rule[-0.175pt]{4.818pt}{0.350pt}}
\put(242,158){\makebox(0,0)[r]{-1e-05}}
\put(1416,158){\rule[-0.175pt]{4.818pt}{0.350pt}}
\put(264,315){\rule[-0.175pt]{4.818pt}{0.350pt}}
\put(242,315){\makebox(0,0)[r]{-5e-06}}
\put(1416,315){\rule[-0.175pt]{4.818pt}{0.350pt}}
\put(264,473){\rule[-0.175pt]{4.818pt}{0.350pt}}
\put(242,473){\makebox(0,0)[r]{0}}
\put(1416,473){\rule[-0.175pt]{4.818pt}{0.350pt}}
\put(264,630){\rule[-0.175pt]{4.818pt}{0.350pt}}
\put(242,630){\makebox(0,0)[r]{5e-06}}
\put(1416,630){\rule[-0.175pt]{4.818pt}{0.350pt}}
\put(264,787){\rule[-0.175pt]{4.818pt}{0.350pt}}
\put(242,787){\makebox(0,0)[r]{1e-05}}
\put(1416,787){\rule[-0.175pt]{4.818pt}{0.350pt}}
\put(264,158){\rule[-0.175pt]{0.350pt}{4.818pt}}
\put(264,113){\makebox(0,0){0}}
\put(264,767){\rule[-0.175pt]{0.350pt}{4.818pt}}
\put(411,158){\rule[-0.175pt]{0.350pt}{4.818pt}}
\put(411,113){\makebox(0,0){0.2}}
\put(411,767){\rule[-0.175pt]{0.350pt}{4.818pt}}
\put(557,158){\rule[-0.175pt]{0.350pt}{4.818pt}}
\put(557,113){\makebox(0,0){0.4}}
\put(557,767){\rule[-0.175pt]{0.350pt}{4.818pt}}
\put(704,158){\rule[-0.175pt]{0.350pt}{4.818pt}}
\put(704,113){\makebox(0,0){0.6}}
\put(704,767){\rule[-0.175pt]{0.350pt}{4.818pt}}
\put(850,158){\rule[-0.175pt]{0.350pt}{4.818pt}}
\put(850,113){\makebox(0,0){0.8}}
\put(850,767){\rule[-0.175pt]{0.350pt}{4.818pt}}
\put(997,158){\rule[-0.175pt]{0.350pt}{4.818pt}}
\put(997,113){\makebox(0,0){1}}
\put(997,767){\rule[-0.175pt]{0.350pt}{4.818pt}}
\put(1143,158){\rule[-0.175pt]{0.350pt}{4.818pt}}
\put(1143,113){\makebox(0,0){1.2}}
\put(1143,767){\rule[-0.175pt]{0.350pt}{4.818pt}}
\put(1290,158){\rule[-0.175pt]{0.350pt}{4.818pt}}
\put(1290,113){\makebox(0,0){1.4}}
\put(1290,767){\rule[-0.175pt]{0.350pt}{4.818pt}}
\put(1436,158){\rule[-0.175pt]{0.350pt}{4.818pt}}
\put(1436,113){\makebox(0,0){1.6}}
\put(1436,767){\rule[-0.175pt]{0.350pt}{4.818pt}}
\put(264,158){\rule[-0.175pt]{282.335pt}{0.350pt}}
\put(1436,158){\rule[-0.175pt]{0.350pt}{151.526pt}}
\put(264,787){\rule[-0.175pt]{282.335pt}{0.350pt}}
\put(45,472){\makebox(0,0)[l]{\shortstack{$Deriv.$\\$Field$\\$EEDSH$}}}
\put(850,68){\makebox(0,0){$r$}}
\put(264,158){\rule[-0.175pt]{0.350pt}{151.526pt}}
\put(264,473){\circle{18}}
\put(268,473){\circle{18}}
\put(281,473){\circle{18}}
\put(325,473){\circle{18}}
\put(425,473){\circle{18}}
\put(554,473){\circle{18}}
\put(689,472){\circle{18}}
\put(840,473){\circle{18}}
\put(996,473){\circle{18}}
\put(1158,472){\circle{18}}
\put(1327,472){\circle{18}}
\put(1363,472){\circle{18}}
\end{picture}
\end{figure}
\begin{figure}[htb]
    % GNUPLOT: LaTeX picture
\setlength{\unitlength}{0.240900pt}
\ifx\plotpoint\undefined\newsavebox{\plotpoint}\fi
\begin{picture}(1500,900)(0,0)
\sbox{\plotpoint}{\rule[-0.175pt]{0.350pt}{0.350pt}}%
\put(264,158){\rule[-0.175pt]{0.350pt}{151.526pt}}
\put(264,158){\rule[-0.175pt]{4.818pt}{0.350pt}}
\put(242,158){\makebox(0,0)[r]{-3.5}}
\put(1416,158){\rule[-0.175pt]{4.818pt}{0.350pt}}
\put(264,248){\rule[-0.175pt]{4.818pt}{0.350pt}}
\put(242,248){\makebox(0,0)[r]{-3}}
\put(1416,248){\rule[-0.175pt]{4.818pt}{0.350pt}}
\put(264,338){\rule[-0.175pt]{4.818pt}{0.350pt}}
\put(242,338){\makebox(0,0)[r]{-2.5}}
\put(1416,338){\rule[-0.175pt]{4.818pt}{0.350pt}}
\put(264,428){\rule[-0.175pt]{4.818pt}{0.350pt}}
\put(242,428){\makebox(0,0)[r]{-2}}
\put(1416,428){\rule[-0.175pt]{4.818pt}{0.350pt}}
\put(264,517){\rule[-0.175pt]{4.818pt}{0.350pt}}
\put(242,517){\makebox(0,0)[r]{-1.5}}
\put(1416,517){\rule[-0.175pt]{4.818pt}{0.350pt}}
\put(264,607){\rule[-0.175pt]{4.818pt}{0.350pt}}
\put(242,607){\makebox(0,0)[r]{-1}}
\put(1416,607){\rule[-0.175pt]{4.818pt}{0.350pt}}
\put(264,697){\rule[-0.175pt]{4.818pt}{0.350pt}}
\put(242,697){\makebox(0,0)[r]{-0.5}}
\put(1416,697){\rule[-0.175pt]{4.818pt}{0.350pt}}
\put(264,787){\rule[-0.175pt]{4.818pt}{0.350pt}}
\put(242,787){\makebox(0,0)[r]{0}}
\put(1416,787){\rule[-0.175pt]{4.818pt}{0.350pt}}
\put(264,158){\rule[-0.175pt]{0.350pt}{4.818pt}}
\put(264,113){\makebox(0,0){0}}
\put(264,767){\rule[-0.175pt]{0.350pt}{4.818pt}}
\put(411,158){\rule[-0.175pt]{0.350pt}{4.818pt}}
\put(411,113){\makebox(0,0){0.2}}
\put(411,767){\rule[-0.175pt]{0.350pt}{4.818pt}}
\put(557,158){\rule[-0.175pt]{0.350pt}{4.818pt}}
\put(557,113){\makebox(0,0){0.4}}
\put(557,767){\rule[-0.175pt]{0.350pt}{4.818pt}}
\put(704,158){\rule[-0.175pt]{0.350pt}{4.818pt}}
\put(704,113){\makebox(0,0){0.6}}
\put(704,767){\rule[-0.175pt]{0.350pt}{4.818pt}}
\put(850,158){\rule[-0.175pt]{0.350pt}{4.818pt}}
\put(850,113){\makebox(0,0){0.8}}
\put(850,767){\rule[-0.175pt]{0.350pt}{4.818pt}}
\put(997,158){\rule[-0.175pt]{0.350pt}{4.818pt}}
\put(997,113){\makebox(0,0){1}}
\put(997,767){\rule[-0.175pt]{0.350pt}{4.818pt}}
\put(1143,158){\rule[-0.175pt]{0.350pt}{4.818pt}}
\put(1143,113){\makebox(0,0){1.2}}
\put(1143,767){\rule[-0.175pt]{0.350pt}{4.818pt}}
\put(1290,158){\rule[-0.175pt]{0.350pt}{4.818pt}}
\put(1290,113){\makebox(0,0){1.4}}
\put(1290,767){\rule[-0.175pt]{0.350pt}{4.818pt}}
\put(1436,158){\rule[-0.175pt]{0.350pt}{4.818pt}}
\put(1436,113){\makebox(0,0){1.6}}
\put(1436,767){\rule[-0.175pt]{0.350pt}{4.818pt}}
\put(264,158){\rule[-0.175pt]{282.335pt}{0.350pt}}
\put(1436,158){\rule[-0.175pt]{0.350pt}{151.526pt}}
\put(264,787){\rule[-0.175pt]{282.335pt}{0.350pt}}
\put(45,472){\makebox(0,0)[l]{\shortstack{$t$}}}
\put(850,68){\makebox(0,0){$r$}}
\put(850,832){\makebox(0,0){$Center \rightarrow Border$}}
\put(264,158){\rule[-0.175pt]{0.350pt}{151.526pt}}
\put(264,787){\circle{18}}
\put(268,783){\circle{18}}
\put(281,770){\circle{18}}
\put(325,726){\circle{18}}
\put(425,638){\circle{18}}
\put(554,540){\circle{18}}
\put(689,457){\circle{18}}
\put(840,381){\circle{18}}
\put(996,320){\circle{18}}
\put(1158,270){\circle{18}}
\put(1327,231){\circle{18}}
\put(1363,224){\circle{18}}
\end{picture}
\end{figure}
\begin{figure}[htb]
    % GNUPLOT: LaTeX picture
\setlength{\unitlength}{0.240900pt}
\ifx\plotpoint\undefined\newsavebox{\plotpoint}\fi
\begin{picture}(1500,900)(0,0)
\sbox{\plotpoint}{\rule[-0.175pt]{0.350pt}{0.350pt}}%
\put(264,158){\rule[-0.175pt]{0.350pt}{151.526pt}}
\put(264,158){\rule[-0.175pt]{4.818pt}{0.350pt}}
\put(242,158){\makebox(0,0)[r]{-3.5}}
\put(1416,158){\rule[-0.175pt]{4.818pt}{0.350pt}}
\put(264,248){\rule[-0.175pt]{4.818pt}{0.350pt}}
\put(242,248){\makebox(0,0)[r]{-3}}
\put(1416,248){\rule[-0.175pt]{4.818pt}{0.350pt}}
\put(264,338){\rule[-0.175pt]{4.818pt}{0.350pt}}
\put(242,338){\makebox(0,0)[r]{-2.5}}
\put(1416,338){\rule[-0.175pt]{4.818pt}{0.350pt}}
\put(264,428){\rule[-0.175pt]{4.818pt}{0.350pt}}
\put(242,428){\makebox(0,0)[r]{-2}}
\put(1416,428){\rule[-0.175pt]{4.818pt}{0.350pt}}
\put(264,517){\rule[-0.175pt]{4.818pt}{0.350pt}}
\put(242,517){\makebox(0,0)[r]{-1.5}}
\put(1416,517){\rule[-0.175pt]{4.818pt}{0.350pt}}
\put(264,607){\rule[-0.175pt]{4.818pt}{0.350pt}}
\put(242,607){\makebox(0,0)[r]{-1}}
\put(1416,607){\rule[-0.175pt]{4.818pt}{0.350pt}}
\put(264,697){\rule[-0.175pt]{4.818pt}{0.350pt}}
\put(242,697){\makebox(0,0)[r]{-0.5}}
\put(1416,697){\rule[-0.175pt]{4.818pt}{0.350pt}}
\put(264,787){\rule[-0.175pt]{4.818pt}{0.350pt}}
\put(242,787){\makebox(0,0)[r]{0}}
\put(1416,787){\rule[-0.175pt]{4.818pt}{0.350pt}}
\put(264,158){\rule[-0.175pt]{0.350pt}{4.818pt}}
\put(264,113){\makebox(0,0){0}}
\put(264,767){\rule[-0.175pt]{0.350pt}{4.818pt}}
\put(411,158){\rule[-0.175pt]{0.350pt}{4.818pt}}
\put(411,113){\makebox(0,0){0.2}}
\put(411,767){\rule[-0.175pt]{0.350pt}{4.818pt}}
\put(557,158){\rule[-0.175pt]{0.350pt}{4.818pt}}
\put(557,113){\makebox(0,0){0.4}}
\put(557,767){\rule[-0.175pt]{0.350pt}{4.818pt}}
\put(704,158){\rule[-0.175pt]{0.350pt}{4.818pt}}
\put(704,113){\makebox(0,0){0.6}}
\put(704,767){\rule[-0.175pt]{0.350pt}{4.818pt}}
\put(850,158){\rule[-0.175pt]{0.350pt}{4.818pt}}
\put(850,113){\makebox(0,0){0.8}}
\put(850,767){\rule[-0.175pt]{0.350pt}{4.818pt}}
\put(997,158){\rule[-0.175pt]{0.350pt}{4.818pt}}
\put(997,113){\makebox(0,0){1}}
\put(997,767){\rule[-0.175pt]{0.350pt}{4.818pt}}
\put(1143,158){\rule[-0.175pt]{0.350pt}{4.818pt}}
\put(1143,113){\makebox(0,0){1.2}}
\put(1143,767){\rule[-0.175pt]{0.350pt}{4.818pt}}
\put(1290,158){\rule[-0.175pt]{0.350pt}{4.818pt}}
\put(1290,113){\makebox(0,0){1.4}}
\put(1290,767){\rule[-0.175pt]{0.350pt}{4.818pt}}
\put(1436,158){\rule[-0.175pt]{0.350pt}{4.818pt}}
\put(1436,113){\makebox(0,0){1.6}}
\put(1436,767){\rule[-0.175pt]{0.350pt}{4.818pt}}
\put(264,158){\rule[-0.175pt]{282.335pt}{0.350pt}}
\put(1436,158){\rule[-0.175pt]{0.350pt}{151.526pt}}
\put(264,787){\rule[-0.175pt]{282.335pt}{0.350pt}}
\put(45,472){\makebox(0,0)[l]{\shortstack{$t$}}}
\put(850,68){\makebox(0,0){$r$}}
\put(850,832){\makebox(0,0){$Border \rightarrow Center$}}
\put(264,158){\rule[-0.175pt]{0.350pt}{151.526pt}}
\put(1363,224){\circle{18}}
\put(1359,225){\circle{18}}
\put(1347,227){\circle{18}}
\put(1300,236){\circle{18}}
\put(1121,281){\circle{18}}
\put(938,341){\circle{18}}
\put(761,419){\circle{18}}
\put(590,516){\circle{18}}
\put(428,635){\circle{18}}
\put(294,757){\circle{18}}
\put(289,761){\circle{18}}
\put(273,778){\circle{18}}
\put(266,785){\circle{18}}
\put(264,787){\circle{18}}
\end{picture}
\end{figure}
\begin{figure}[htb]
    % GNUPLOT: LaTeX picture
\setlength{\unitlength}{0.240900pt}
\ifx\plotpoint\undefined\newsavebox{\plotpoint}\fi
\begin{picture}(1500,900)(0,0)
\sbox{\plotpoint}{\rule[-0.175pt]{0.350pt}{0.350pt}}%
\put(264,158){\rule[-0.175pt]{282.335pt}{0.350pt}}
\put(264,158){\rule[-0.175pt]{0.350pt}{151.526pt}}
\put(264,158){\rule[-0.175pt]{4.818pt}{0.350pt}}
\put(242,158){\makebox(0,0)[r]{0}}
\put(1416,158){\rule[-0.175pt]{4.818pt}{0.350pt}}
\put(264,228){\rule[-0.175pt]{4.818pt}{0.350pt}}
\put(242,228){\makebox(0,0)[r]{0.1}}
\put(1416,228){\rule[-0.175pt]{4.818pt}{0.350pt}}
\put(264,298){\rule[-0.175pt]{4.818pt}{0.350pt}}
\put(242,298){\makebox(0,0)[r]{0.2}}
\put(1416,298){\rule[-0.175pt]{4.818pt}{0.350pt}}
\put(264,368){\rule[-0.175pt]{4.818pt}{0.350pt}}
\put(242,368){\makebox(0,0)[r]{0.3}}
\put(1416,368){\rule[-0.175pt]{4.818pt}{0.350pt}}
\put(264,438){\rule[-0.175pt]{4.818pt}{0.350pt}}
\put(242,438){\makebox(0,0)[r]{0.4}}
\put(1416,438){\rule[-0.175pt]{4.818pt}{0.350pt}}
\put(264,507){\rule[-0.175pt]{4.818pt}{0.350pt}}
\put(242,507){\makebox(0,0)[r]{0.5}}
\put(1416,507){\rule[-0.175pt]{4.818pt}{0.350pt}}
\put(264,577){\rule[-0.175pt]{4.818pt}{0.350pt}}
\put(242,577){\makebox(0,0)[r]{0.6}}
\put(1416,577){\rule[-0.175pt]{4.818pt}{0.350pt}}
\put(264,647){\rule[-0.175pt]{4.818pt}{0.350pt}}
\put(242,647){\makebox(0,0)[r]{0.7}}
\put(1416,647){\rule[-0.175pt]{4.818pt}{0.350pt}}
\put(264,717){\rule[-0.175pt]{4.818pt}{0.350pt}}
\put(242,717){\makebox(0,0)[r]{0.8}}
\put(1416,717){\rule[-0.175pt]{4.818pt}{0.350pt}}
\put(264,787){\rule[-0.175pt]{4.818pt}{0.350pt}}
\put(242,787){\makebox(0,0)[r]{0.9}}
\put(1416,787){\rule[-0.175pt]{4.818pt}{0.350pt}}
\put(264,158){\rule[-0.175pt]{0.350pt}{4.818pt}}
\put(264,113){\makebox(0,0){0}}
\put(264,767){\rule[-0.175pt]{0.350pt}{4.818pt}}
\put(411,158){\rule[-0.175pt]{0.350pt}{4.818pt}}
\put(411,113){\makebox(0,0){0.2}}
\put(411,767){\rule[-0.175pt]{0.350pt}{4.818pt}}
\put(557,158){\rule[-0.175pt]{0.350pt}{4.818pt}}
\put(557,113){\makebox(0,0){0.4}}
\put(557,767){\rule[-0.175pt]{0.350pt}{4.818pt}}
\put(704,158){\rule[-0.175pt]{0.350pt}{4.818pt}}
\put(704,113){\makebox(0,0){0.6}}
\put(704,767){\rule[-0.175pt]{0.350pt}{4.818pt}}
\put(850,158){\rule[-0.175pt]{0.350pt}{4.818pt}}
\put(850,113){\makebox(0,0){0.8}}
\put(850,767){\rule[-0.175pt]{0.350pt}{4.818pt}}
\put(997,158){\rule[-0.175pt]{0.350pt}{4.818pt}}
\put(997,113){\makebox(0,0){1}}
\put(997,767){\rule[-0.175pt]{0.350pt}{4.818pt}}
\put(1143,158){\rule[-0.175pt]{0.350pt}{4.818pt}}
\put(1143,113){\makebox(0,0){1.2}}
\put(1143,767){\rule[-0.175pt]{0.350pt}{4.818pt}}
\put(1290,158){\rule[-0.175pt]{0.350pt}{4.818pt}}
\put(1290,113){\makebox(0,0){1.4}}
\put(1290,767){\rule[-0.175pt]{0.350pt}{4.818pt}}
\put(1436,158){\rule[-0.175pt]{0.350pt}{4.818pt}}
\put(1436,113){\makebox(0,0){1.6}}
\put(1436,767){\rule[-0.175pt]{0.350pt}{4.818pt}}
\put(264,158){\rule[-0.175pt]{282.335pt}{0.350pt}}
\put(1436,158){\rule[-0.175pt]{0.350pt}{151.526pt}}
\put(264,787){\rule[-0.175pt]{282.335pt}{0.350pt}}
\put(45,472){\makebox(0,0)[l]{\shortstack{$I$}}}
\put(850,68){\makebox(0,0){$r$}}
\put(850,832){\makebox(0,0){$Center \rightarrow Border$}}
\put(264,158){\rule[-0.175pt]{0.350pt}{151.526pt}}
\put(264,158){\circle{18}}
\put(268,162){\circle{18}}
\put(281,174){\circle{18}}
\put(325,217){\circle{18}}
\put(425,303){\circle{18}}
\put(554,399){\circle{18}}
\put(689,483){\circle{18}}
\put(840,560){\circle{18}}
\put(996,624){\circle{18}}
\put(1158,678){\circle{18}}
\put(1327,722){\circle{18}}
\put(1363,730){\circle{18}}
\end{picture}
\end{figure}
\begin{figure}[htb]
    % GNUPLOT: LaTeX picture
\setlength{\unitlength}{0.240900pt}
\ifx\plotpoint\undefined\newsavebox{\plotpoint}\fi
\begin{picture}(1500,900)(0,0)
\sbox{\plotpoint}{\rule[-0.175pt]{0.350pt}{0.350pt}}%
\put(264,158){\rule[-0.175pt]{282.335pt}{0.350pt}}
\put(264,158){\rule[-0.175pt]{0.350pt}{151.526pt}}
\put(264,158){\rule[-0.175pt]{4.818pt}{0.350pt}}
\put(242,158){\makebox(0,0)[r]{0}}
\put(1416,158){\rule[-0.175pt]{4.818pt}{0.350pt}}
\put(264,228){\rule[-0.175pt]{4.818pt}{0.350pt}}
\put(242,228){\makebox(0,0)[r]{0.1}}
\put(1416,228){\rule[-0.175pt]{4.818pt}{0.350pt}}
\put(264,298){\rule[-0.175pt]{4.818pt}{0.350pt}}
\put(242,298){\makebox(0,0)[r]{0.2}}
\put(1416,298){\rule[-0.175pt]{4.818pt}{0.350pt}}
\put(264,368){\rule[-0.175pt]{4.818pt}{0.350pt}}
\put(242,368){\makebox(0,0)[r]{0.3}}
\put(1416,368){\rule[-0.175pt]{4.818pt}{0.350pt}}
\put(264,438){\rule[-0.175pt]{4.818pt}{0.350pt}}
\put(242,438){\makebox(0,0)[r]{0.4}}
\put(1416,438){\rule[-0.175pt]{4.818pt}{0.350pt}}
\put(264,507){\rule[-0.175pt]{4.818pt}{0.350pt}}
\put(242,507){\makebox(0,0)[r]{0.5}}
\put(1416,507){\rule[-0.175pt]{4.818pt}{0.350pt}}
\put(264,577){\rule[-0.175pt]{4.818pt}{0.350pt}}
\put(242,577){\makebox(0,0)[r]{0.6}}
\put(1416,577){\rule[-0.175pt]{4.818pt}{0.350pt}}
\put(264,647){\rule[-0.175pt]{4.818pt}{0.350pt}}
\put(242,647){\makebox(0,0)[r]{0.7}}
\put(1416,647){\rule[-0.175pt]{4.818pt}{0.350pt}}
\put(264,717){\rule[-0.175pt]{4.818pt}{0.350pt}}
\put(242,717){\makebox(0,0)[r]{0.8}}
\put(1416,717){\rule[-0.175pt]{4.818pt}{0.350pt}}
\put(264,787){\rule[-0.175pt]{4.818pt}{0.350pt}}
\put(242,787){\makebox(0,0)[r]{0.9}}
\put(1416,787){\rule[-0.175pt]{4.818pt}{0.350pt}}
\put(264,158){\rule[-0.175pt]{0.350pt}{4.818pt}}
\put(264,113){\makebox(0,0){0}}
\put(264,767){\rule[-0.175pt]{0.350pt}{4.818pt}}
\put(411,158){\rule[-0.175pt]{0.350pt}{4.818pt}}
\put(411,113){\makebox(0,0){0.2}}
\put(411,767){\rule[-0.175pt]{0.350pt}{4.818pt}}
\put(557,158){\rule[-0.175pt]{0.350pt}{4.818pt}}
\put(557,113){\makebox(0,0){0.4}}
\put(557,767){\rule[-0.175pt]{0.350pt}{4.818pt}}
\put(704,158){\rule[-0.175pt]{0.350pt}{4.818pt}}
\put(704,113){\makebox(0,0){0.6}}
\put(704,767){\rule[-0.175pt]{0.350pt}{4.818pt}}
\put(850,158){\rule[-0.175pt]{0.350pt}{4.818pt}}
\put(850,113){\makebox(0,0){0.8}}
\put(850,767){\rule[-0.175pt]{0.350pt}{4.818pt}}
\put(997,158){\rule[-0.175pt]{0.350pt}{4.818pt}}
\put(997,113){\makebox(0,0){1}}
\put(997,767){\rule[-0.175pt]{0.350pt}{4.818pt}}
\put(1143,158){\rule[-0.175pt]{0.350pt}{4.818pt}}
\put(1143,113){\makebox(0,0){1.2}}
\put(1143,767){\rule[-0.175pt]{0.350pt}{4.818pt}}
\put(1290,158){\rule[-0.175pt]{0.350pt}{4.818pt}}
\put(1290,113){\makebox(0,0){1.4}}
\put(1290,767){\rule[-0.175pt]{0.350pt}{4.818pt}}
\put(1436,158){\rule[-0.175pt]{0.350pt}{4.818pt}}
\put(1436,113){\makebox(0,0){1.6}}
\put(1436,767){\rule[-0.175pt]{0.350pt}{4.818pt}}
\put(264,158){\rule[-0.175pt]{282.335pt}{0.350pt}}
\put(1436,158){\rule[-0.175pt]{0.350pt}{151.526pt}}
\put(264,787){\rule[-0.175pt]{282.335pt}{0.350pt}}
\put(45,472){\makebox(0,0)[l]{\shortstack{$I$}}}
\put(850,68){\makebox(0,0){$r$}}
\put(850,832){\makebox(0,0){$Border \rightarrow Center$}}
\put(264,158){\rule[-0.175pt]{0.350pt}{151.526pt}}
\put(1363,730){\circle{18}}
\put(1359,730){\circle{18}}
\put(1347,727){\circle{18}}
\put(1300,716){\circle{18}}
\put(1121,667){\circle{18}}
\put(938,602){\circle{18}}
\put(761,522){\circle{18}}
\put(590,423){\circle{18}}
\put(428,306){\circle{18}}
\put(294,187){\circle{18}}
\put(289,183){\circle{18}}
\put(273,167){\circle{18}}
\put(266,160){\circle{18}}
\put(264,158){\circle{18}}
\end{picture}
\end{figure}

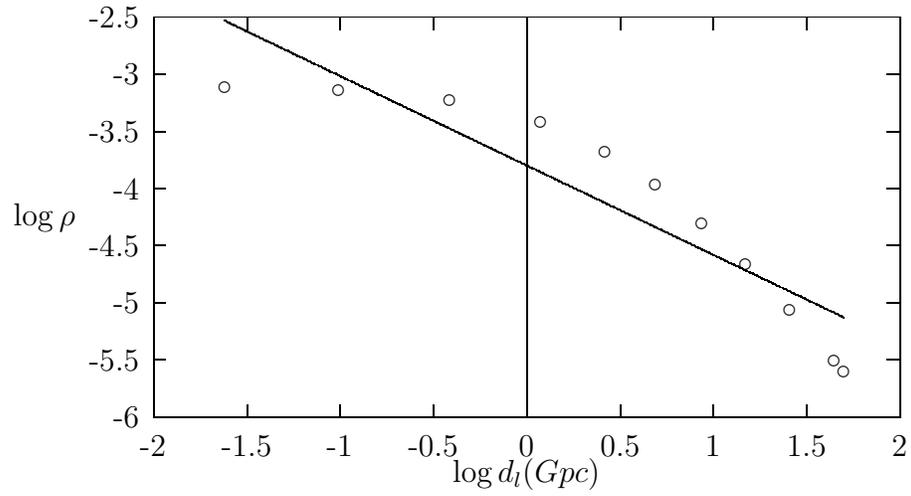
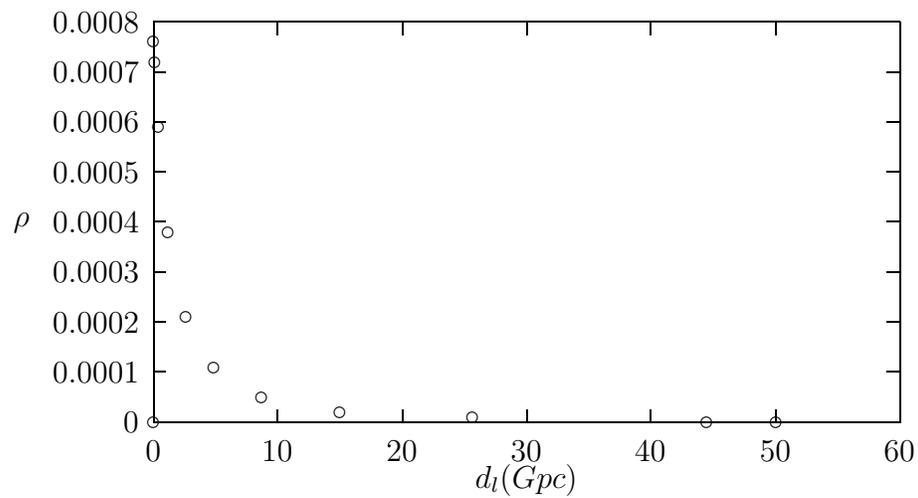
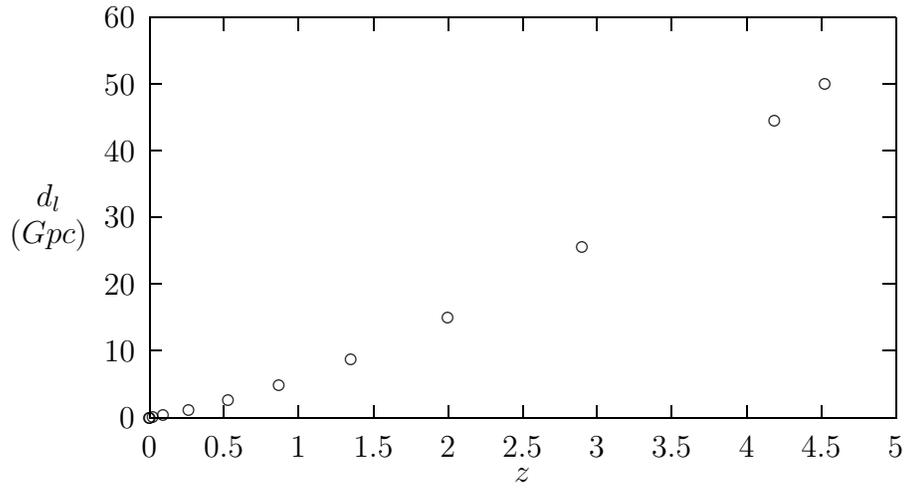
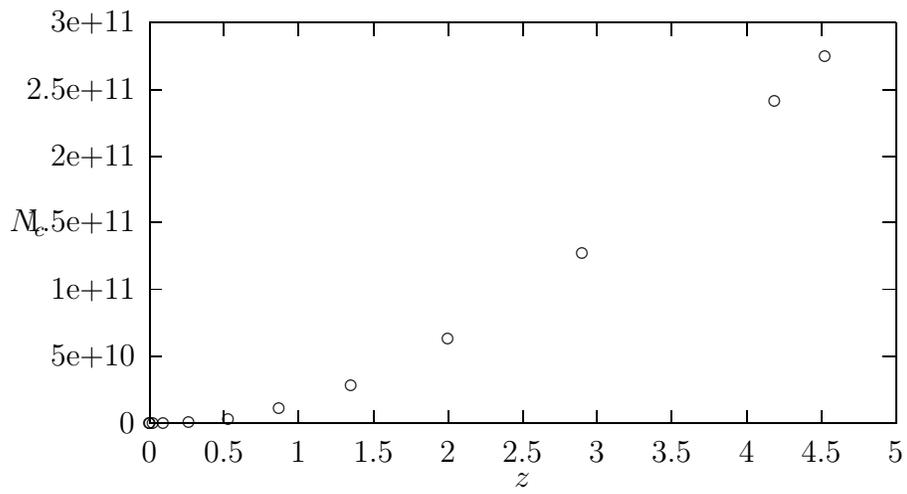
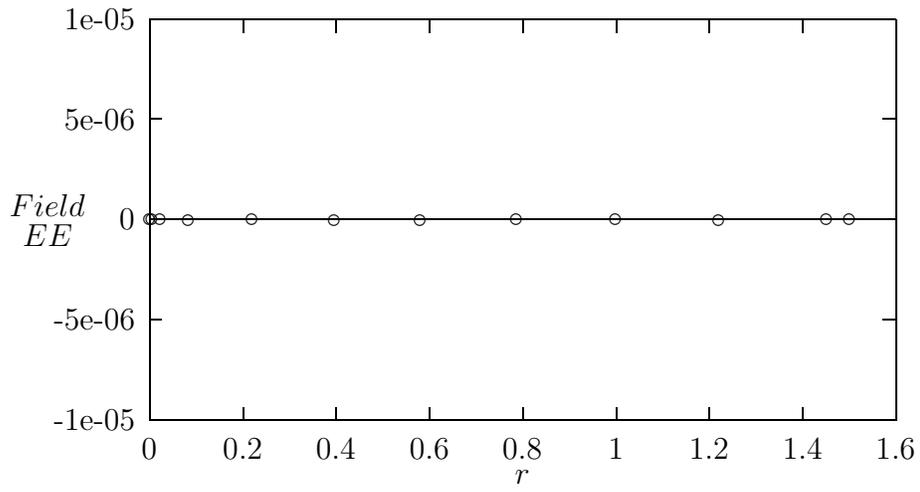
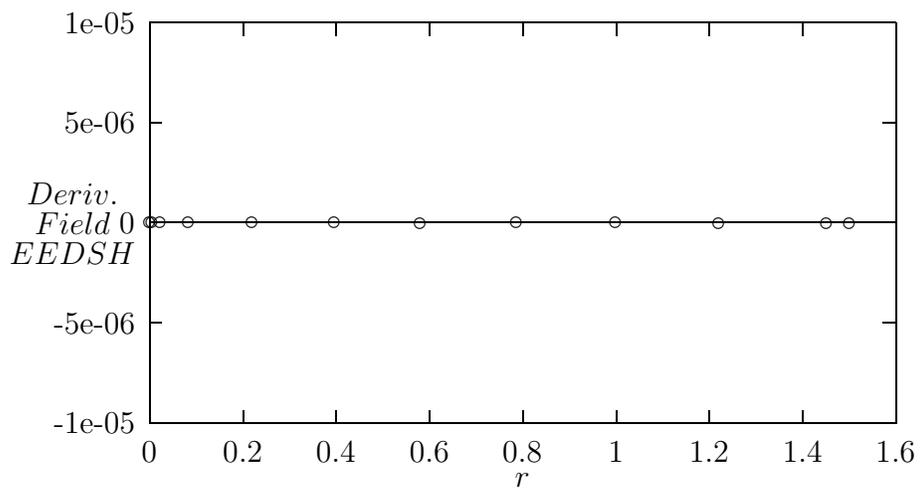
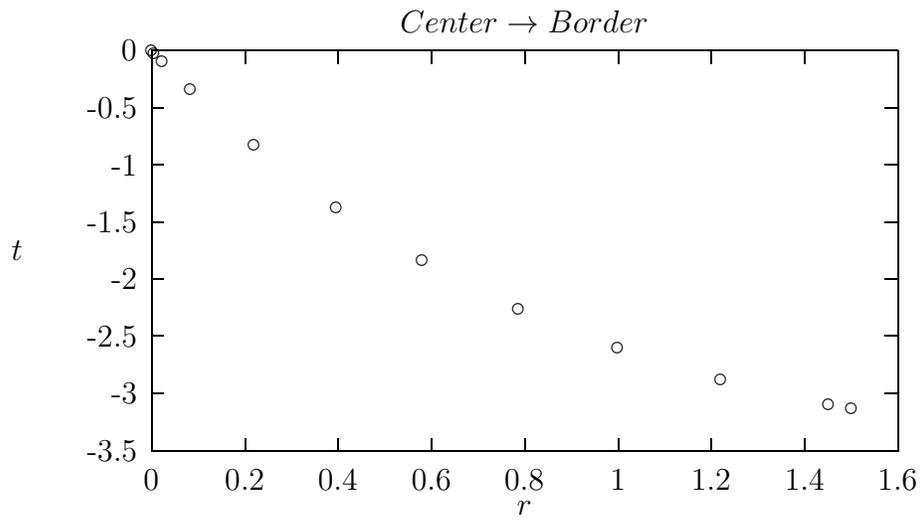
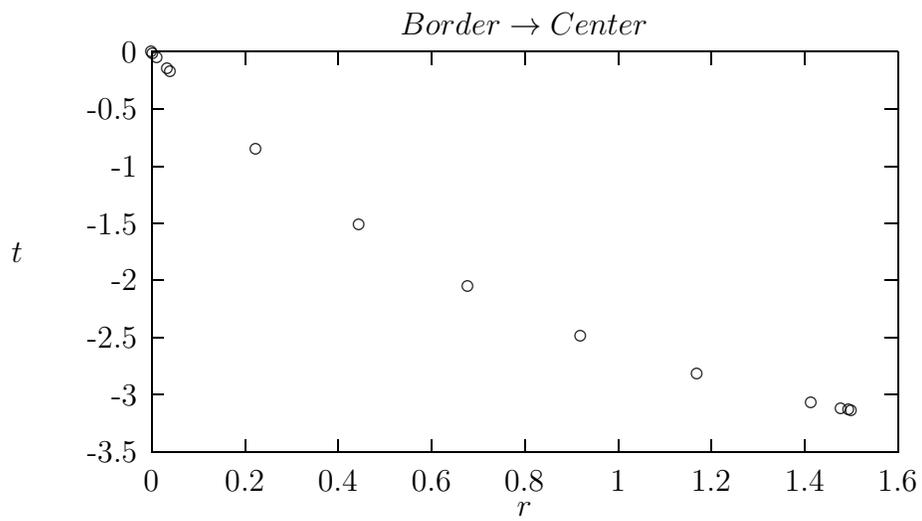
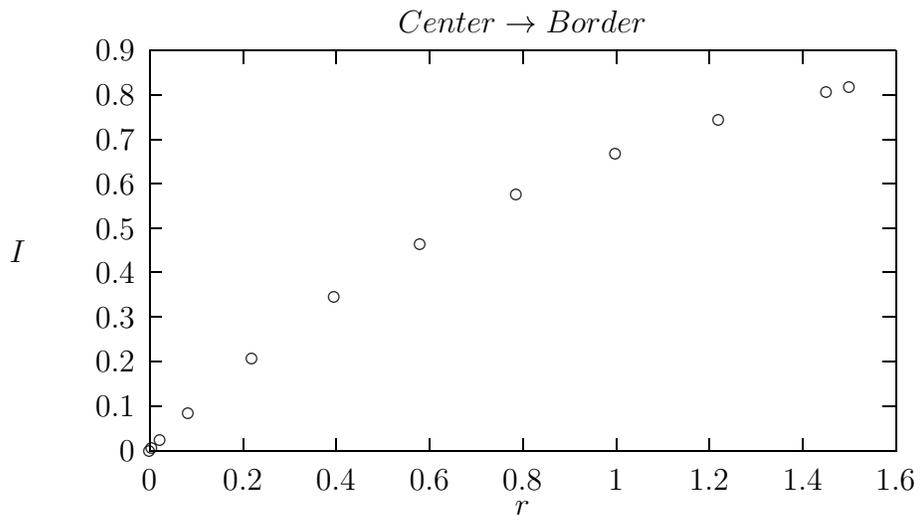
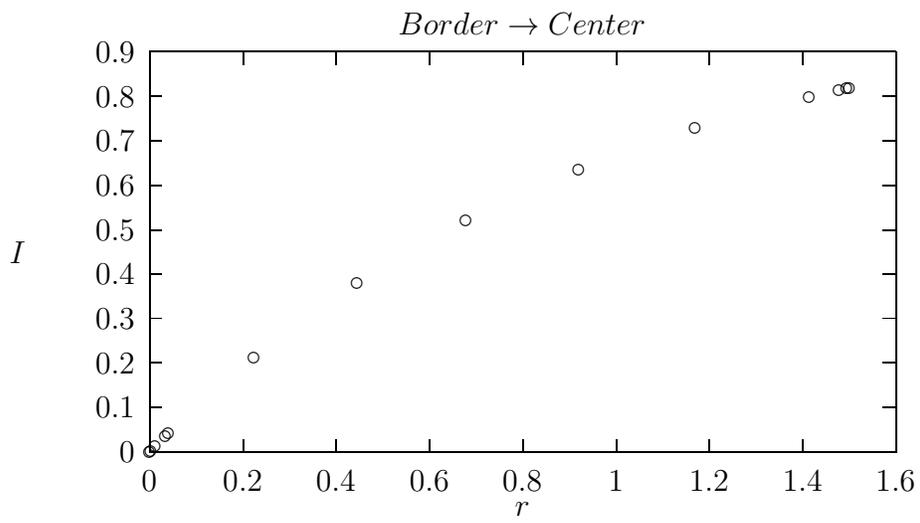
\end{document}